# La$_{1-x}$Mn$_{1-y}$O$_{3\pm\delta}$ buffer layers on inclined substrate deposited MgO templates for coated conductors

Oleksiy Troshyn[1,3], Christian Hoffmann[2], Veit Große[3], Jens Hänisch[4], Lucas Becker[3], and Rudolf Gross[1,5]

[1] Physik-Department, Technische Universität München, 85748 Garching, Germany
[2] Ceraco Ceramic Coating GmbH, Rote-Kreuz-Str. 8, 85737 Ismaning, Germany
[3] THEVA Dünnschichttechnik GmbH, Rote-Kreuz-Str. 8, 85737 Ismaning, Germany
[4] Institute for Technical Physics, Karlsruhe Institute of Technology (KIT), PO Box 36 40, 76021 Karlsruhe, Germany
[5] Walther-Meißner-Institut, Bayerische Akademie der Wissenschaften, 85748 Garching, Germany
E-mail: oleksiy.troshyn@tum.de

## Abstract

Most commercial high-temperature superconducting coated conductors based on ion beam assisted MgO deposited templates use LaMnO$_3$ (LMO) films as the terminating buffer layer. In contrast, coated conductors based on inclined substrate deposition (ISD)-MgO technology are still produced with homoepitaxial (homoepi)-MgO as the cap layer. In this work we report on the deposition of LMO buffer layers on ISD-MgO/homoepi-MgO by electron beam physical vapor deposition. The growth parameters of textured LMO films were studied systematically and their properties were optimized regarding the critical current densitiy ($J_c$) of the subsequently deposited DyBa$_2$Cu$_3$O$_{7-\delta}$ (DyBCO) superconducting films. LMO films without outgrowths at the surface were obtained at growth rates of up to 4 Å/s. Despite the formation of non-stoichiometric LMO films containing 59 % La, single-phase films were obtained at substrate temperatures below 775 °C and at oxygen partial pressures of up to 4×10$^{-4}$ mbar due to a large homogeneity region towards La. The $J_c$ values of DyBCO films deposited on LMO were found to be independent of the LMO thickness in a range from 50 nm to 450 nm. DyBCO films on LMO reach $J_c$ = 0.83 MA/cm$^2$ at 77 K in zero applied field. This value is up to 30 % higher than those of DyBCO films grown directly on homoepi-MgO. The wide range of LMO growth parameters and higher $J_c$ values of DyBCO on LMO compared to DyBCO on homoepi-MgO make this material attractive for its use in manufacturing coated conductors based on ISD-MgO technology.

## 1. Introduction

Coated conductor (CC) tapes based on REBa$_2$Cu$_3$O$_{7-\delta}$ (RE rare earth) (REBCO) high temperature superconductors (HTS) have unique properties at low temperatures that can be used in electrical power transmission cables, high-field magnets and rotating electrical machines [1], [2]. The enabling property for such applications is the high critical current density ($J_c$) of these REBCO films, which primarily can be achieved through epitaxial growth on biaxially textured buffer layers. The most popular technique for texture creation, which is used by the majority of manufacturing companies of HTS-CC, is the ion beam assisted MgO deposition (IBAD-MgO) of the template [3]. An alternative texturing technique is the inclined substrate deposition (ISD) which also generates a biaxially textured MgO layer [4], [5]. For the manufacturing of CC tapes, the ISD-MgO technique is attractive due to its simplicity and the high deposition rates that can be achieved. However, the mismatch of the lattice parameters between the MgO template and the REBCO layer, which is usually grown directly on top of it, can decrease the achieveable critical current density [6].

In order to reduce this lattice mismatch and therefore increase $J_c$, an additional interlayer can be deposited between REBCO and MgO, regardless of the texturing method [7]. SrRuO$_3$ [8], SrTiO$_3$ [9], [10], LaMnO$_3$ (LMO) [11], [12] and CeO$_2$ [13] have been investigated as materials for such a final buffer layer on the IBAD architecture using e.g. pulsed laser depositon (PLD) and rf-sputtering. On ISD-MgO templates, so far SrRuO$_3$ [14], [15] and YSZ/CeO$_2$ [16], [17] films were deposited as cap layers via PLD, whereas only LMO or LMO/CeO$_2$ architectures are used on IBAD templates in commercial CC manufacturing. Commercial tapes based on the ISD technology are still produced without any additional cap layer [18]. According to Xiong et al. [19], LMO not only yields high $J_c$ values but also allows for a large window of deposition parameters. This ensures a stable and reliable film growth, which is crucial when choosing a suitable cap layer. Until today there are only a few publications describing the parameter boundaries for the LMO growth and their effect on *RE*BCO properties [11], [12]. In particular there is no prior work that provides information on the deposition of the LMO buffer layer on the ISD-MgO template and on the evaporation of an LMO granulate by an electron beam.

For the first time we demonstrate here the epitaxial growth of LMO films on a template consisting of ISD-MgO covered with a homoepitaxial (homoepi)-MgO layer by the electron beam physical vapor deposition (EB-PVD). We discuss the morphology and the texture of the fabricated LMO layers depending on their thickness and deposition parameters. We report the successful growth of DyBCO films on such ISD-MgO templates with LMO cap layer. At optimized deposition parameters of LMO, an increase of $J_c$ by up to 30% is achieved for DyBCO films grown on LMO compared to DyBCO films grown on homoepi-MgO layers.

**2. Experimental procedure**

*2.1 ISD MgO templates*

Biaxially aligned MgO templates, which consist of ISD-MgO/homoepi-MgO layers deposited on 12 mm wide Hastelloy C276 metal tape, were used as substrates. Details of the substrate preparation and the ISD-MgO texturing technique have been published previously [20]. ISD-MgO and homoepi-MgO layers have been deposited on a continuously processed tape with optimized growth conditions. All substrates used in the experiments were cut from the same tape with uniform MgO texture. This ensures the elimination of any influence of the substrate on the properties of the layers that were deposited on top of it. The in-plane and out-of-plane texture values of ISD-MgO (200) are 9.5° and 4.5°, respectively.

*2.2 LMO buffer layers*

The LMO buffer layers were deposited on the MgO templates by evaporating an LMO granulate with a grain size of < 0.5 mm by means of EB-PVD. The deposition system and the measurement of the substrate temperature ($T_{sub}$) are described in detail in Ref. [21].

For LMO film growth the substrate temperature and the oxygen pressure were varied from 500 °C to 900 °C and from $3\times10^{-5}$ mbar to $1.1\times10^{-3}$ mbar respectively. The LMO film thickness was varied in the range from 50 nm to 450 nm and the average growth rate was varied from 2 Å/s to 5 Å/s.

*2.3 DyBCO films*

The DyBCO films were deposited on LMO as well as directly on MgO buffer layers by evaporating a DyBCO granulate with a grain size of < 0.5 mm by EB-PVD. Unlike for the LMO deposition, the substrates were placed on a rotating substrate holder. By rotating the substrates continuously between a deposition zone with a pressure of $2\times10^{-4}$ mbar and an oxygen pocket with an oxygen pressure of about $5\times10^{-3}$ mbar at a frequency of 5 Hz, a subsequent vapor condensation and oxidation take place during film growth. The substrates were radiation-heated to 700 °C for the DyBCO deposition. More details of the setup can be found in [22]. The evaporation rate was measured by a quartz crystal monitor and was kept at 3 Å/s. The thickness of the DyBCO films was 1 µm.

The deposition of DyBCO films was used as a quick test to qualify the LMO buffer layers. A readily available DyBCO granulate, not optimized regarding the stoichiometry yielding the highest $J_c$ values, of Dy:Ba:Cu = 20:25:55 was used for this purpose. It is known that $J_c$ values of DyBCO films depend strongly on their composition [23] and can sharply decrease outside the optimal composition range [24], [25], [21]. Therefore, in the following discussion it has to be kept in mind that in the present work the DyBCO composition was not carefully optimized for yielding the highest $J_c$.

*2.4 Sample characterization*

The surface morphology and the thickness of the samples were analyzed by scanning electron microscopy (SEM) using a Hitachi SU8000. The composition of the film constituents was analyzed by energy dispersive x-ray spectroscopy (EDX), performed on an FEI Titan transmission electron microscope (TEM). Specimens for TEM were prepared by focused ion beam. The sample stoichiometry was measured by inductively coupled plasma atomic emission spectroscopy (ICP). As Hastelloy C276 contains manganese, the use of the metal substrate would introduce a significant error in the measurement of the LMO film composition. Hence, LMO films designated for ICP measurements were deposited on MgO single crystals. It was assumed that the layer stoichiometry does not significantly depend on the substrate material and its temperature. The structural properties of the LMO and DyBCO films were characterized by X-ray diffraction (XRD) using a Philips X'Pert X-Ray Diffractometer with Cu-Kα radiation. The texture was qualified using X-ray ω-rocking curves of the (200) LMO reflection. The full width at half maximum (FWHM) of the rocking curve peak is noted as Δω. The critical temperature ($T_c$) and critical current density values of the DyBCO samples were measured inductively (77 K, self-field) with a setup described in [26].

## 3. Results and discussion

### 3.1 Deposition rate

A basic parameter in thin film deposition is the deposition rate. Therefore we first studied the dependence of the film properties on the deposition rate by growing 100 nm thick LMO films at average evaporation rates of 2 Å/s and 5 Å/s on tapes with ISD-MgO/homoepi-MgO architecture (Fig. 1).

The surface morphology of the LMO films deposited with the rate of 2 Å/s is similar to that of the homoepi-MgO layer. It consists of terraces that are aligned in the same way as the MgO surface. Every terrace has a smooth surface with sharp edges and corners. Such terraces are also visible at an average deposition rate of 5 Å/s, but their surfaces are disturbed by the presence of outgrowths, which form predominantely within the terraces.

Notable during the deposition were occasional short-term rate bursts of up to 4 Å/s at an average rate of 2 Å/s, and beyond 10 Å/s at an average rate of 5 Å/s. These rate bursts originate from the formation of large droplets of the melt in the evaporation zone and their subsequent sudden evaporation. Such droplet formation is common for the EB-PVD method and appears to an increasing degree with higher evaporation rates, as more grains are fed into the evaporation zone.

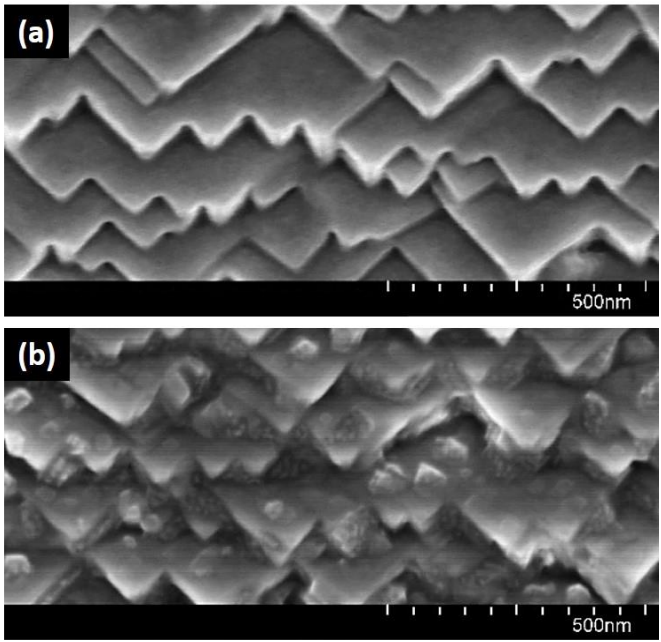

Figure 1. Surface morphology of 100 nm thick LMO films deposited at $T_{sub}$ = 750 °C and $p(O_2)$ = 4×10⁻⁴ mbar. The average deposition rate is (a) 2 Å/s and (b) 5 Å/s. In contrast to (a) where the layer surface is smooth, outgrowths can be found in (b).

Apparently, the rate bursts beyond 10 Å/s are the limiting factor for a smooth surface morphology rather than the average rate of 5 Å/s. As an LMO layer with many outgrowths offers a poor base for the epitaxial growth of REBCO, the LMO films described in the following were produced at an average deposition rate of 2 Å/s to ensure that the inevitable rate bursts are below the critical value.

### 3.2 Film stoichiometry

When stoichiometric LMO granulate is evaporated during the EB-PVD process, the deposited LMO films have an excess of La of about 9 % compared to the composition of the original granulate (Fig. 2). The precursor stoichiometry is therefore not exactly transferred into the deposited film by using EB-PVD. This is caused by the decomposition of the LMO powder and the formation of La- and Mn-rich droplets during evaporation. The evaporation characteristic of these different droplets is varying, resulting in the observed film stoichiometry [27].

The film stoichiometry was found to be reproducible with an accuracy of 1.5 %. The observed variation in stoichiometry is mainly caused by an incomplete evaporation of the granulate. In case of good thermal coupling, droplets containing La and Mn can coagulate on the water cooled copper crucible and the EB power then is too low to evaporate these droplets. They are visible as residues on the crucible after deposition. In addition, the variation in stoichiometry transfer can also arise due to non-optimized electron beam settings and the granulate track geometry, since an occasional splashing of droplets from the evaporation zone was observed during the deposition.

In order to deposit stoichiometric LMO films, a granulate with $MnO_z$ excess has been evaporated. As a result, films with the surface morphology shown in Fig. 1b were obtained due to the formation and rapid evaporation of large droplets from the melt, leading to evaporation rates beyond 10 Å/s. Thus, an increased manganese content in the granulate affected the stability of the evaporation rate. Since the observed effect could not be eliminated, the stoichiometric LMO granulate was used in all depositions described below.

### 3.3 Substrate temperature

In order to find the optimal temperature range for the growth of high-quality LMO buffer layers suitable for the subsequent REBCO deposition, the substrate temperature was varied from

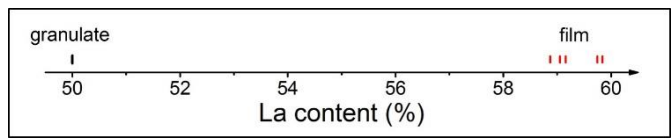

Figure 2. La content in the evaporated LMO granulate and in a series of five different deposited films (measured by ICP analysis). The reproducibility of the film stoichiometry is in the range of 1.5 %.

500 °C to 900 °C and its influence on morphology and crystallinity was studied. For simplicity only the samples deposited at an oxygen partial pressure of $4\times10^{-4}$ mbar are described here, since a preliminary parameter screening showed that at this pressure high-quality LMO buffer layers can be grown.

The 100 nm thick LMO films grown on the homoepi-MgO layer were found to be epitaxially textured in the entire investigated temperature range, which is evident from the intense ($h$00) LMO peaks in the ω-2θ diffractograms (Fig. 3) and sharp pole figures (not shown here) of these films. The FWHM of the LMO (200) rocking curve, Δω, decreases (inset in Fig. 3) with growing temperature, indicating a steady improvement of the film crystallinity. The improved texturing/alignment can be explained by an increased surface diffusion at higher temperatures. The adsorbed atoms then have a sufficiently high energy to reach and occupy the most energetically favourable positions on the crystal surface, i.e. kink and step sites. Hence, the number of defects in the LMO crystal structure decreases. Nevertheless, high substrate temperatures also favour the growth of misoriented crystallites corresponding to the LMO (111) reflection.

The surfaces of all LMO films consist of terraces, thus replicating the MgO underlayer over a wide temperature range (Fig. 4). However, pronounced differences in the layer morphology can be distinguished at a closer inspection. For instance, an increase in substrate temperature from 500 °C to 700 °C leads to smoother terrace surfaces with sharper edges and corners. They can be explained by an increased surface diffusion of deposited atoms at higher temperatures. In the temperature range from 700 °C to 775 °C the morphology of the films does not change noticeably. At 800 °C bright spots can be observed in the SEM images of the layer surface. With a further temperature increase to 900 °C, the size of these defects increases and the sharp edges and corners of the terraces fade.

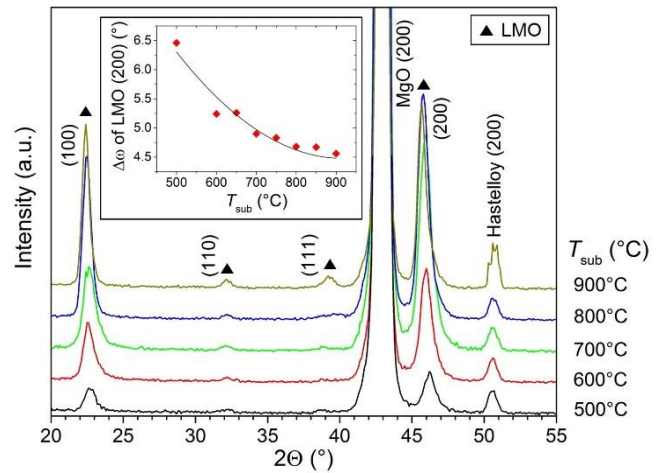

Figure 3. ω-2θ diffractograms of LMO films grown at different substrate temperatures. The insert shows the rocking curve width Δω of the (200) LMO reflections. All LMO layers were deposited at $p(O_2) = 4\times10^{-4}$ mbar.

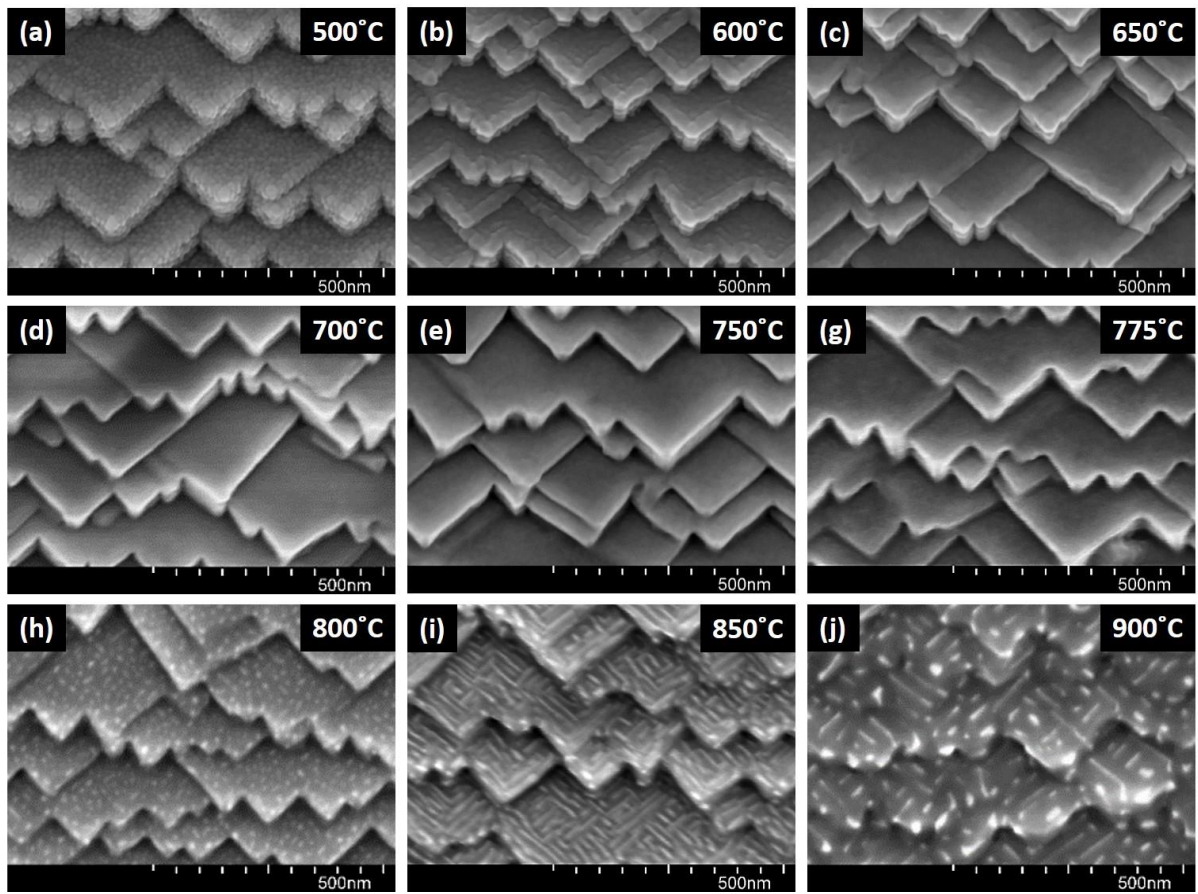

Figure 4. Surface morphology of 100 nm thick LMO films grown in the temperature range from 500 °C to 900 °C.

High resolution TEM EDX elemental maps of LMO films deposited at $T_{sub}$ = 700 °C and $T_{sub}$ = 850 °C, accordingly without and with bright spots visible in SEM images in Fig. 4, were taken to analyze the composition of the precipitates, which correspond to these bright spots (Fig. 5). While La and Mn atoms are homogeneously distributed in the film grown at 700 °C (Fig. 5a), the film in Fig. 5b ($T_{sub}$ = 850 °C) obviously contains precipitates with a La excess and a Mn deficiency. They are evenly distributed across the entire film thickness and have likely been formed during the film growth. Thus, the bright spots in the SEM images of films deposited at temperatures above 800 °C correspond to La-rich precipitates.

The absence of these precipitates at low substrate temperatures, their formation at high temperatures and their increasing size with increasing substrate temperature can be explained using the phase diagram of ½La$_2$O$_3$-MnO$_z$ shown in Fig. 6. It was calculated in [28] in air, i.e. at an oxygen partial pressure of 210 mbar, and is used as a rough guideline since it can notably change at the LMO deposition conditions ($p(O_2) \leq 1.1 \times 10^{-3}$ mbar) [29]. In the La-Mn-O system a congruently melting, non-stoichiometric perovskite $(La_{1-d}Mn_d)_{1-x}Mn_{1-y}O_{3\pm\delta}$ [29] with a wide homogeneity range extending mainly in the La direction is formed. For a La-rich sample, as in our case, $d \approx 0$. The red arrow marks the path for a temperature rise in a La-rich LMO compound with 56 % La. According to the phase diagram, only a single phase, i.e. the perovskite, forms for such a composition at temperatures below 760 °C (region ①). At about 760 °C, the La$_2$O$_3$ phase is crystallized from the perovskite at the two-phase boundary (point ②). With growing temperature, the amount of La$_2$O$_3$ increases at the expense of perovskite according to the lever rule, thus changing the composition of the perovskite (region ③).

The described sequence of the phase separation in the phase diagram with increasing temperature is in good agreement with our experimental observations. The La-rich inclusions appear in LMO films with 59 % La at $T_{sub}$ = 800 °C and grow with increasing substrate temperature. Therefore, the observed precipitates can be attributed to the La$_2$O$_3$ phase. In contrast to the phase diagram, we observe a wider homogeneity region towards La, which is probably a result of different experimental conditions.

The following experiment confirms the assumption of the phase separation in the LMO film with increasing substrate temperature. An LMO film with a La content of 59 %, deposited at 700 °C and $4 \times 10^{-4}$ mbar oxygen, was annealed for 30 minutes at $T_{sub}$ = 900 °C and $p(O_2) = 4 \times 10^{-4}$ mbar and then quickly cooled down to room temperature. While initially no precipitates were visible on the surface (Fig. 7a), bright spots appeared after the annealing procedure mainly at the terrace boundaries (Fig. 7b). In a further annealing step up to 900 °C, these precipitates could not be dissolved by a slow cool-down at an oxygen pressure of 210 mbar. Therefore, an irreversible reaction with the formation of the $La_2O_3$ phase has taken place at high temperatures. According to this experiment, the ½$La_2O_3$-$MnO_z$ phase diagram can be used as a scheme to describe the growth of the single-phase LMO films at $p(O_2) \leq 1.1\times10^{-3}$ mbar.

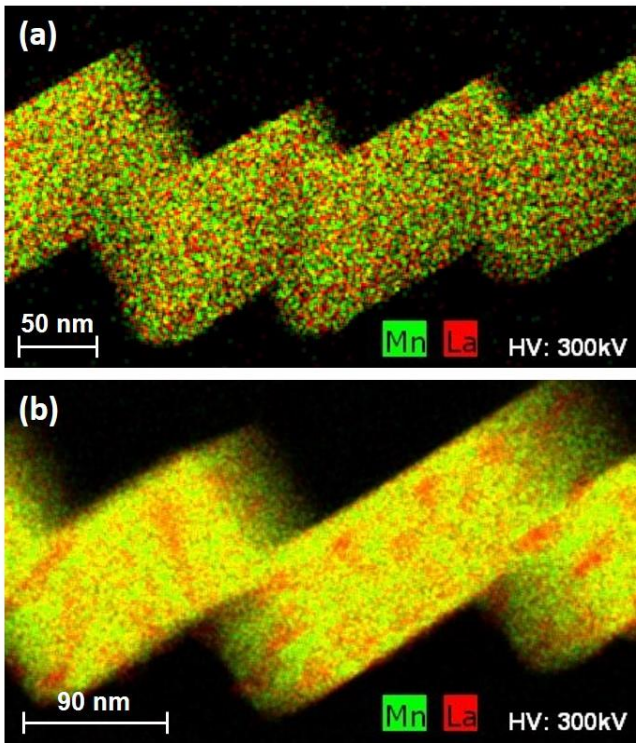

Figure 5. High resolution TEM EDX of LMO films deposited at a substrate temperature of (a) 700 °C and (b) 850 °C. In (b) the formation of precipitates with a La excess and a Mn deficiency is revealed.

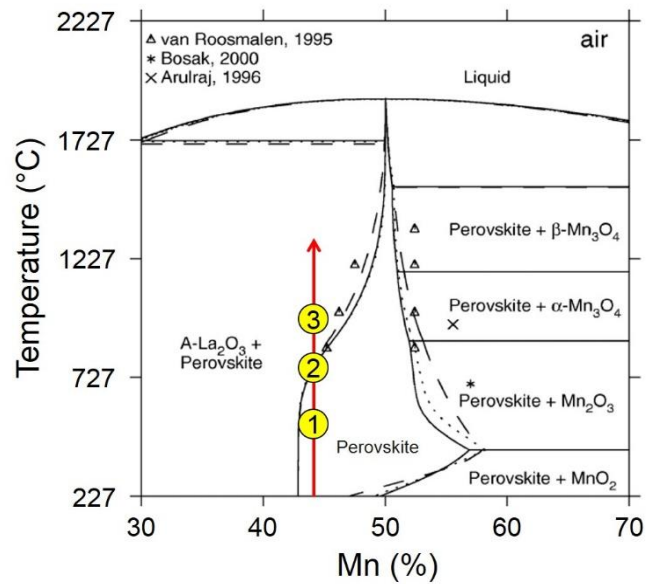

Figure 6. The phase diagram of ½ La$_2$O$_3$ - MnO$_z$ calculated in air in [28]. Point 2 marks the boundary between two different phase regions, labelled with numbers 1 and 3.

The critical current density of the DyBCO films on LMO changes significantly with LMO growth temperature (Fig. 8). In particular the $J_c$ value of DyBCO increases with the LMO deposition temperature until it reaches a maximum at $T_{sub}$ = 700 °C. On further increasing the temperature to 800 °C, $J_c$ decreases slightly and then sharply drops at $T_{sub}$ > 800 °C. The observed $J_c$ distribution strongly correlates with the surface roughness of the LMO buffer layers. We determined the surface roughness from SEM images (cf. Fig. 4) since it is not possible to measure it with an atomic force microscope due to the morphology of the film surface consisting of inclined terraces. The LMO surface is rough at temperatures less than 650 °C due to low surface diffusion of adsorbed particles, and at temperatures above 800 °C due to the formation of the $La_2O_3$ precipitates. Though the $La_2O_3$ precipitates form already at 800 °C, they do not seem to disturb the DyBCO growth, probably due to their size being still small.

In summary, LMO buffer layers for the subsequent growth of DyBCO (or in general $RE$BCO) films with a high critical current density can be deposited in a wide temperature range from 650 °C to 800 °C despite the excess of La. Such a large temperature window for the deposition process makes LMO an attractive candidate for application in the production of HTS-CC.

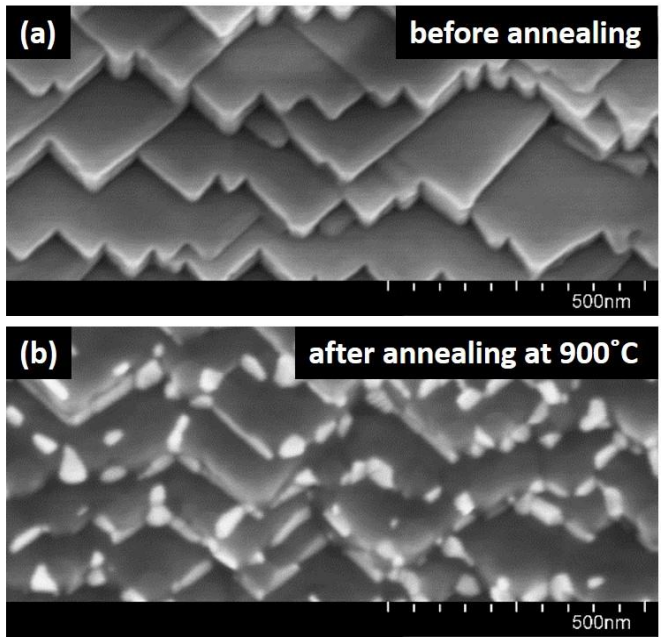

Figure 7. (a) Surface morphology of the 100 nm thick LMO film deposited at $T_{sub}$ = 700 °C and $p(O_2)$ = 4×10$^{-4}$ mbar. (b) Morphology of the same sample after annealing for 30 minutes at $T_{sub}$ = 900 °C and $p(O_2)$ = 4×10$^{-4}$ mbar.

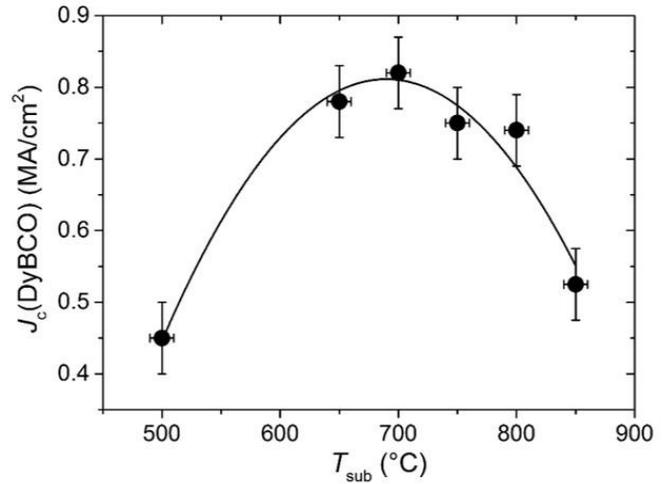

Figure 8. Critical current density $J_c$ of 1 μm thick DyBCO films grown on LMO buffer layers plotted versus the substrate temperature during LMO deposition. All LMO layers were grown at $p(O_2)$ = 4×10$^{-4}$ mbar.

### 3.4 Oxygen partial pressure

In Refs. [27] and [30], it has been shown that during EB-PVD evaporation of both $RE$BCO and MgO a vaporization of the constituent oxides, with their partial decomposition into metal and oxygen atoms, occurs. These metal atoms and oxides are adsorbed on the substrate surface and consequently additional oxygen is required for the formation of a fully oxidized epitaxial film. It is reasonable to assume that this is also the case during LMO evaporation. Both manganese and lanthanum atoms and oxides are evaporated and adsorbed on the sample. The additional oxygen is provided by the background oxygen atmosphere in the vacuum chamber.

Unexpectedly crystalline LMO films grow textured on MgO even at extremely low oxygen partial pressures and thus in the entire investigated range from $p(O_2) = 3\times10^{-5}$ mbar to $p(O_2) = 1.1\times10^{-3}$ mbar. This becomes evident from the intense ($h$00) LMO reflections in the diffractogram (Fig. 9a). The formation of the LMO phase even at the low pressure of $p(O_2) = 3\times10^{-5}$ mbar implies that either particles containing lanthanum and manganese mainly vaporize as oxides, or that crystalline LMO films can also be formed at a significant oxygen deficiency as reported in [31], [32]. When studying the impact of oxygen partial pressure on the growth of LMO buffer layers, their quality is again evaluated regarding the $J_c$ values achieved for the subsequently deposited DyBCO superconducting films.

As shown in Fig. 9b, the highest critical current densities were achieved for DyBCO films on LMO buffers deposited at oxygen pressures of $p(O_2) \leq 4\times10^{-4}$ mbar and at temperatures from 650 °C to approx. 750 °C. The achieved $J_c$ values on the LMO buffer layers with optimized process parameters are consistently higher than for a reference DyBCO sample grown directly on an MgO template. Possible reasons for this observation are discussed in the next paragraph. With increasing temperature and/or oxygen pressure beyond this optimal process parameter window the $J_c$ values decrease significantly.

The decrease in quality of DyBCO films on LMO layers that were deposited at temperatures above 800 °C is not surprising. The formation of $La_2O_3$ precipitates at these temperatures and their impact was already discussed above. However, such crystallites are formed already at a substrate temperature of 750 °C and an oxygen pressure of $p(O_2) = 1.1 \times 10^{-3}$ mbar, as can be seen in the corresponding SEM image in Fig. 10. Furthermore, an increase of oxygen pressure intensifies the formation of the $La_2O_3$ crystallites with regard to their density and size, as clearly seen on the samples deposited at 850 °C. This result suggests a shrinkage of the La solubility range in the perovskite under oxidizing conditions similar to increasing temperature (Fig. 6). The dependence of the Mn solubility in LMO on oxygen partial pressures has been studied in [32], [33]. However, for La it is reported here to our knowledge for the first time. Interestingly, both the number of $La_2O_3$ precipitates in SEM images and the intensity of the (110) LMO peak in the $\omega$-$2\theta$ diffractogram increase with increasing oxygen pressure (Fig. 9a), although the in-plane and out-of-plane texture of LMO does not depend on $p(O_2)$ and is 8.9° and 5.0°, respectively. Since the intensity of the (110) peak is proportional to the number of misoriented grains in the LMO buffer layer, the $J_c$ values of the DyBCO films deposited on top of them are expected to decrease, which is in agreement with the experimental observation.

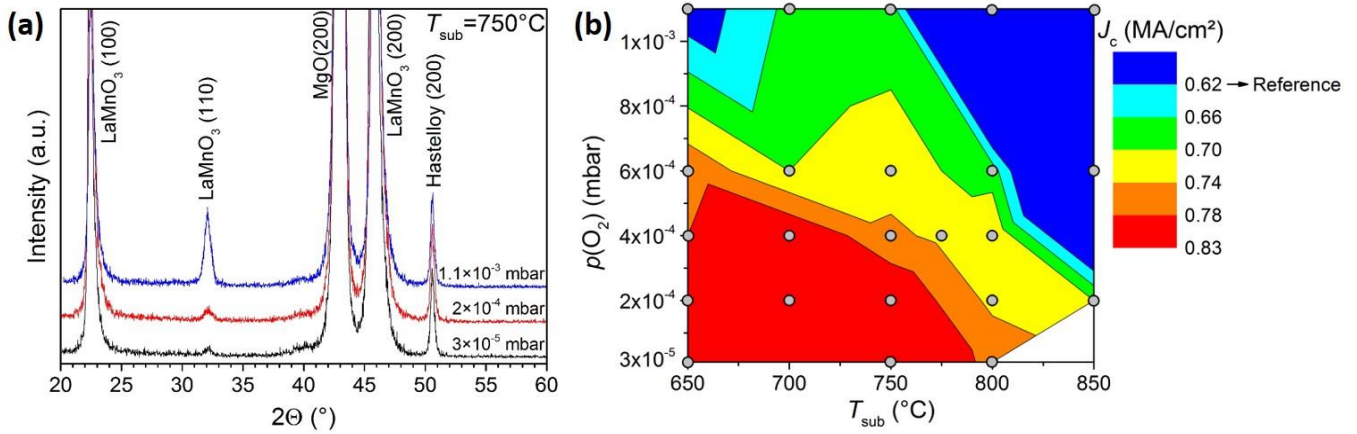

Figure 9. (a) ω-2θ diffractograms of LMO films grown at $T_{sub}$ = 750 °C. The increasing intensity of the (110) LMO peak with increasing oxygen partial pressure indicates the favoured growth of the misoriented grains. (b) Critical current density $J_c$ of DyBCO films deposited on 100 nm thick LMO buffer layers at varying oxygen partial pressures and substrate temperatures.

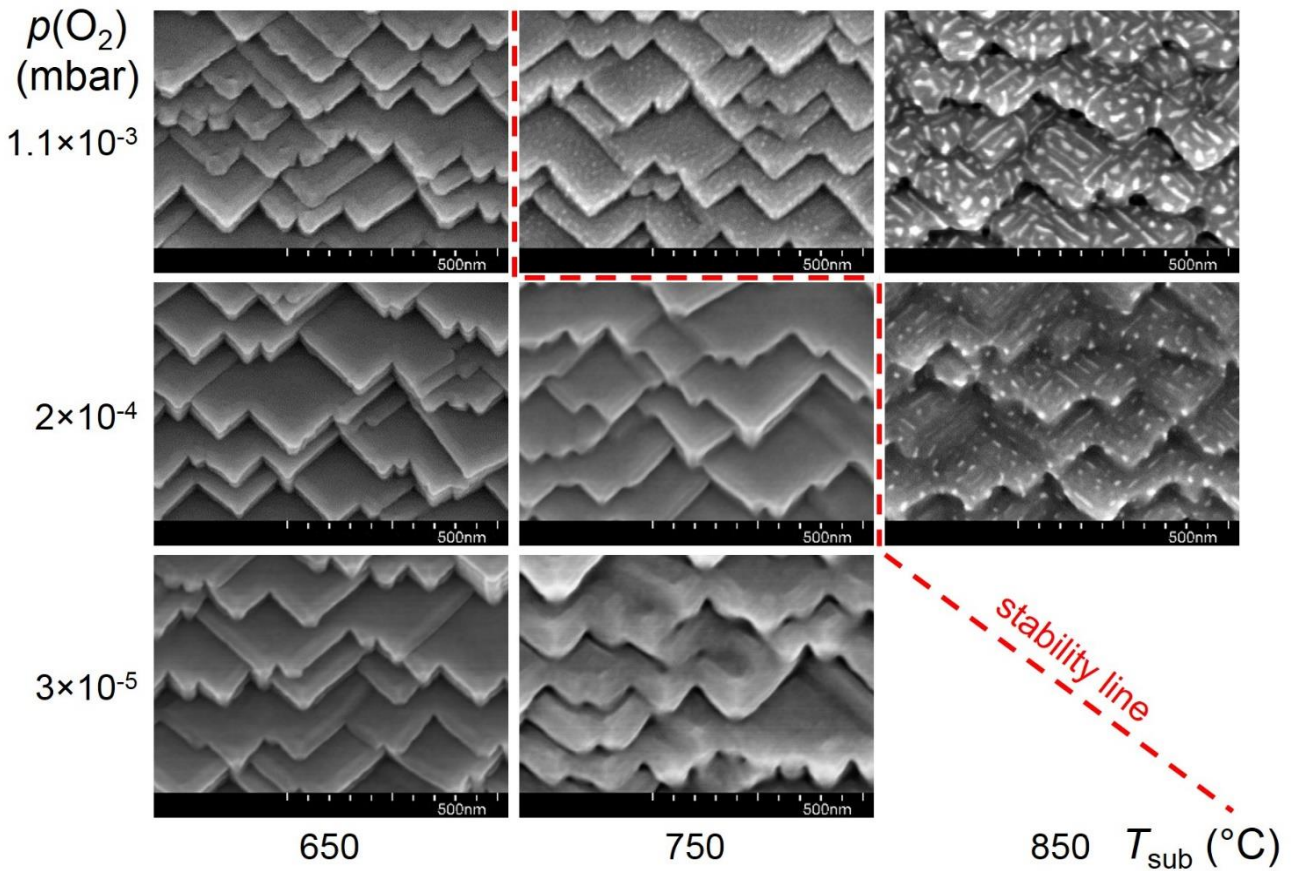

Figure 10. Surface morphology of 100 nm thick LMO buffer layers deposited at various temperatures and oxygen partial pressures. The red dotted line indicates the stability line of the LMO phase. An increase of the oxygen partial pressure and/or the substrate temperature favours the formation of $La_2O_3$ precipitates.

### 3.5 Film thickness

So far our results indicate that the critical current density of DyBCO layers can be improved by more than 30 % using LMO cap layers with a thickness of 100 nm. In order to determine the minimal and maximal LMO layer thickness, still allowing for an increase in $J_c$, we investigated the $J_c$ values of DyBCO films on LMO buffer layer with a thickness between 50 nm and 450 nm. The LMO was deposited at $p(O_2) = 2\times10^{-4}$ mbar and $T_{sub} = 700$ °C.

The surface morphology of the 50 nm thick film has sharp edges and smooth terraces (Fig. 11a). The terraces and the steps of MgO are completely covered by LMO with homogeneously distributed thickness (Fig. 11b), a coverage which could not be clearly seen at smaller thicknesses. In contrast, for a 300 nm thick film the uniform and flat alignment of terraces on the surface is no longer maintained and the edges of the terraces are curved or show facets, which can be seen in both SEM top view and cross-section of this layer (see Fig. 11c and d). The surface area of the terraces with a smooth surface is therefore decreased.

The origin of the development of the observed surface morphology with increasing film thickness is not exactly known. However, there is evidence for a change of growth conditions with increasing thickness indicated by an increasing blackening of thick layers. While a 50 nm thick LMO film still appears white or transparent, the films thicker than 300 nm look dark grey. This colour change can affect the sample temperature at a constant heating power due to radiant heating of only the rear side of the substrate. A change of the emission coefficient in the Stefan-Boltzmann law, associated with the surface blackening, leads to a higher radiation loss of the film with growing thickness and thus a decreasing substrate temperature. The resulting reduction of the surface diffusion length of the adsorbed particles might lead to the observed development of the film morphology.

The $J_c$ and $T_c$ values of DyBCO on LMO are independent of the LMO thickness in the studied range (Fig. 12). It implies, that the observed morphology change of LMO films with increasing thickness does not have a strong influence on the superconducting properties of DyBCO. The absence of any dependence of $J_c$ of REBCO films on the LMO buffer layer thickness in the range from 30 nm to 240 nm was also found for DyBCO films grown on IBAD-MgO templates [12].

The $J_c$ and $T_c$ values of DyBCO films on LMO buffers are higher compared to the reference sample, DyBCO grown on a homoepi-MgO layer. We found no correlation between $J_c$ and the in-plane texture of DyBCO (measured at the (103) peak) on these buffer layers. Hence there should be other reasons for this observation. There may be a different strain state of DyBCO on LMO due to a smaller lattice mismatch of about 1 % between these two materials compared to 8 % between DyBCO and MgO [11]. As a result, a larger stoichiometry window might be expected when growing DyBCO on LMO compared to homoepi-MgO. Or there may be a slight Mg poisoning of DyBCO grown on MgO, which reduces $T_c$ [34]. Meanwhile, Mn possibly diffusing into DyBCO grown on LMO would not poison it [35]. Also assuming a similar $J_c(T/T_c)$ dependence for all these samples, higher $T_c$ values easily explain higher $J_c$ values.

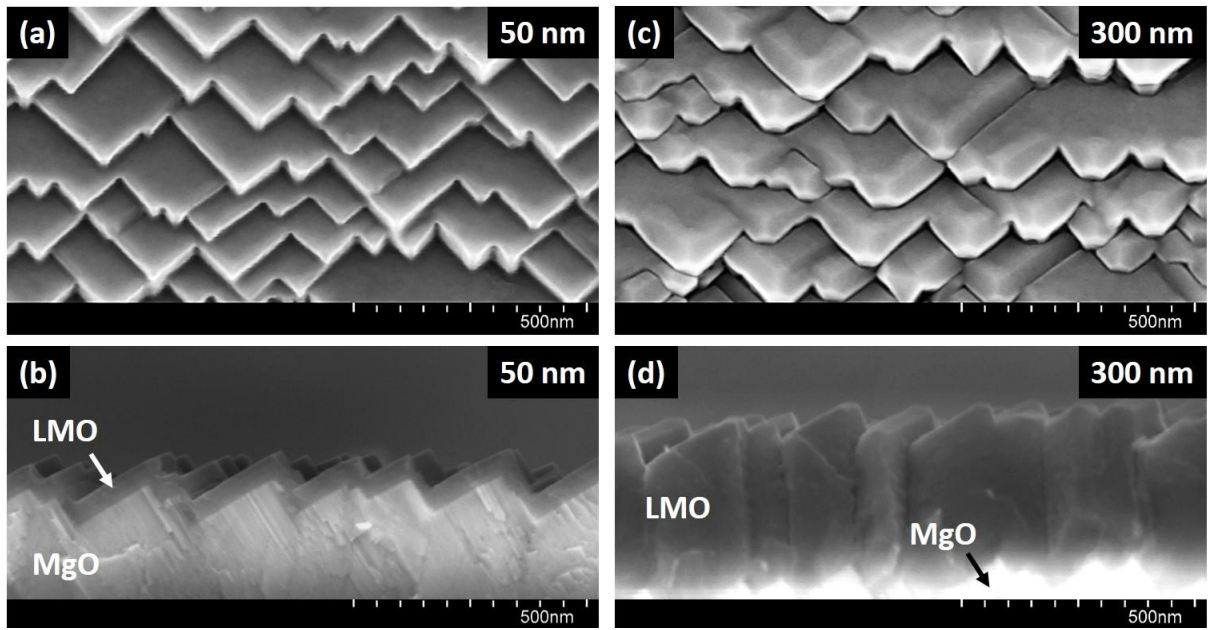

*Figure 11. Surface morphology (a, c) and SEM cross-sections (b, d) of the 50 nm and the 300 nm thick LMO layers deposited at $T_{sub} = 700$ °C and $p(O_2) = 2\times10^{-4}$ mbar on homoepi-MgO. The MgO terraces are fully covered with LMO already at layer thickness of 50 nm. The sharp edges of the 50 nm thick film are shown in contrast to the rounded edges of the 300 nm thick film.*

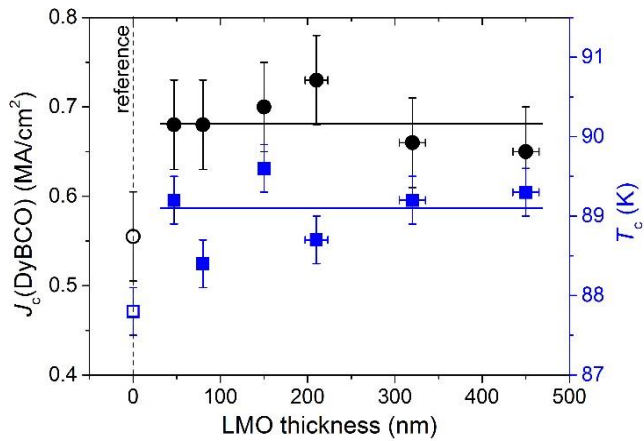

*Figure 12. $J_c$ and $T_c$ of DyBCO films plotted versus the thickness of the underlying LMO buffer layers. For comparison, $J_c$ and $T_c$ values of a reference sample with a DyBCO film grown on a homoepi-MgO layer are shown.*

We note that the highest $J_c$ values of DyBCO, grown on LMO and homoepi-MgO in this work are lower than the state-of-the-art $J_c$ values, which for DyBCO on ISD-MgO/homoepi-MgO reach $J_c \approx 2$ MA/cm$^2$ [23], [36]. Taking into account an optimized ISD-texture, similar to [23], and further assuming that relevant deposition parameters of DyBCO such as substrate temperature, oxygen pressure, and deposition rate were close to optimal, we believe that the low $J_c$ is caused by the non-optimized DyBCO composition and does not represent the limit for these buffer layers. Nevertheless, we were able to optimize the LMO growth parameters regarding the change of the critical current density of the subsequently deposited DyBCO films.

## 4. Conclusions

For the first time, LMO buffer layers were deposited on ISD-MgO templates using EB-PVD. LMO films with 59 % La are formed when evaporating a stoichiometric LMO granulate by an electron beam. Despite this LMO films grow epitaxially on MgO in a very wide range of temperatures from 500 °C to 900 °C and partial oxygen pressures from $3 \times 10^{-5}$ mbar to $1.1 \times 10^{-3}$ mbar. Due to non-stoichiometry of LMO and its large homogeneity region towards La, single-phase films were obtained at temperatures $T_{sub} \leq 775$ °C and pressures $p(O_2) \leq 4 \times 10^{-4}$ mbar. The limitations of LMO growth with EB-PVD include the formation of outgrowths at the film surface at evaporation rates above approx. 5 Å/s as well as misorientations at high temperatures and oxygen partial pressures. In the latter case the LMO misorientations are correlated to the formation of La$_2$O$_3$ precipitates due to La excess in the samples and may be eliminated by a deposition of stoichiometric films.

DyBCO films grown on LMO buffer layers with thickness avobe 50 nm have up to 30 % higher self-field $J_c$ values at 77 K compared to DyBCO grown directly on the homoepi-MgO layer. This shows that LMO buffer layer grown on top of ISD-MgO template is a promising candidate for usage in industrially produced HTS-CC.

## Acknowledgements

The authors thank colleagues at THEVA Dünnschichttechnik GmbH for technical assistance and W. Prusseit for supporting this research.